# Constructing A Small Strain Potential for Multi-Scale Modeling

A. Mallik, K. Runge, H-P Cheng, and J. Dufty
*Department of Physics, University of Florida,*
*Gainesville, FL-32611*

**Abstract**

Simulation of bulk materials with some components locally far from equilibrium usually requires a computationally intensive quantum mechanical description to capture the relevant mechanisms (e.g. effects of chemistry). Multi-scale modeling entails a compromise whereby the most accurate quantum description is used only where needed and the remaining bulk of the material is replaced by a simpler classical system of point particles. The problem of constructing an appropriate potential energy function for this classical system is addressed here. For problems relating to fracture, a consistent embedding of a quantum (QM) domain in its classical (CM) environment requires that the classical system should yield the same structure and elastic properties as the QM domain for states near equilibrium. It is proposed that an appropriate classical potential can be constructed using *ab initio* data on the equilibrium structure and weakly strained configurations calculated from the quantum description, rather than the more usual approach of fitting to a wide range of empirical data. This scheme is illustrated in detail for a model system, a silica nanorod that has the proper stiochiometric ratio of Si:O as observed in real silica. The potential energy is chosen to be pairwise additive, with the same pair potential functional form as familiar phenomenological TTAM potential. Here, the parameters are determined using a genetic algorithm with force data obtained directly from a quantum calculation. The resulting potential gives excellent agreement with properties of the reference quantum calculations both for structure (bond lengths, bond angles) and elasticity (Young's modulus). The proposed method for constructing the classical potential is carried out for two different choices for the quantum mechanical description: a transfer Hamiltonian method (NDDO with coupled-cluster parameterization) and density functional theory (with plane wave basis set and PBE exchange correlation functional). The quality of the potentials obtained in both cases is quite good, although the two quantum rods have significant differences.

## I. Introduction

For studying phenomena like crack propagation, stress corrosion etc. one can apply *ab initio* quantum mechanics only to the small reactive regions because these methods are computationally too intensive for application to the entire bulk. Instead, the remainder of the material is typically described by more efficient but less accurate classical molecular dynamics (MD) methods. This scheme of combining different methods for different regions in a sample is known as multi-scale modeling.

One of challenges with such multi-scale modeling is the compatibility of the quantum (QM) domain with the classical (CM) region. Independent individual treatment of the two regions can result in a mismatch of many physical quantities like forces, charge densities, and elastic constants across the QM/CM boundary. To avoid this, the interatomic forces used in the MD simulation must be chosen to preserve selected properties of the underlying quantum theory used in the QM domain. The process of constructing such a potential is illustrated here for a simple test system, providing the prerequisite for a consistent embedding of a QM domain in a classical MD region with a seamless coupling between the two regions. This classical potential is developed using *ab initio* data on the equilibrium structure and weakly strained configurations calculated from the quantum description, rather than the more usual approach of fitting to a wide range of empirical data. The latter is appropriate when a classical description of the entire material is reasonable (e.g., near equilibrium) so that maximum correlation between the material simulated and the real material is attained. In contrast, for multi-scale modeling of strain to failure the primary constraints on the potential are those of internal theoretical consistency between the quantum and classical descriptions. The final accuracy of the model is therefore set by the quality of the quantum description, not by the fitting of the potential to experiments. Further comment on this point is given in the Conclusions section.

To illustrate this approach and to allow a quantitative test of its accuracy a relatively small system is considered for which the chosen quantum method can be applied globally as well as to the partitioned QM/CM model. Selected properties (e.g., force data) are first calculated using a suitable approximate quantum mechanical method that incorporates the relevant physical mechanisms. Next, a parameterized functional form is chosen for the classical potential energy function. Here, it is taken to be pairwise additive so that only a small set of pair potential parameters must be determined. These are chosen to be of the same functional form as current phenomenological models, but with the parameters chosen differently. An initial choice is made using a genetic algorithm to give a good fit to the selected quantum properties. The algorithm effectively locates a domain of the relevant parametric minimum for the given constraints; the parameters found are then checked for stability, followed by a global scaling (described in Appendix), resulting in the final choice for the parameters and determining all pair potentials.

The test system is a silica nanorod [1] with the proper stoichiometric ratio of Si:O observed in real silica. For the multi-scale modeling of this rod (Figure 1.) one of the rings near the center of the nanorod is treated quantum mechanically and the rest is treated classically. The method of embedding a QM domain in its CM environment is done by replacing the atoms at the boundary of the QM and CM domains by pseudo-atoms and approximating the rest of the rod by dipoles.

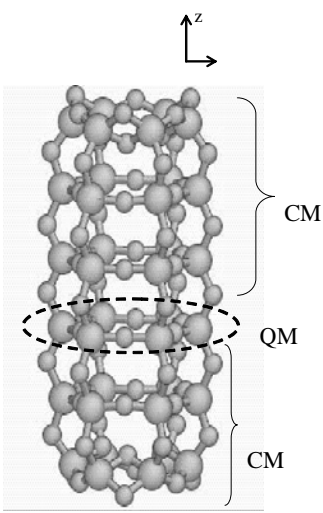

Figure 1. Front view of the nanorod showing partitioning into QM and CM domains

The details of this embedding method have been described elsewhere [2], and will not be considered further here. Instead the following focuses on the choice of the classical potential energy and the resulting properties of the CM region. Specifically, the objective is to determine the potential energy such that a CM simulation of the entire nanorod is indistinguishable from the QM description for all relevant properties in states near equilibrium. This is a prerequisite for expecting that the multi-scale modeling above can be internally consistent.

The standard classical pair potentials for silica, TTAM [3] and BKS [4], fail to satisfy these criteria. Although these potentials have been parameterized using data for equilibrium states, this is not sufficient to yield low strain behavior consistent with the transfer Hamiltonian (TH) as measured by the value of elastic constants. Figure 2 shows the stress-strain curve for this silica nanorod using the standard TTAM and BKS pair potentials (see Eq. 1 below). Also shown for comparison are the reference results obtained from one of the two quantum methods considered here (TH [5]) for the entire rod. These curves were generated by MD simulation for the entire nanorod subjected to longitudinal strain along the z direction by pulling the end caps with a fixed velocity. The remaining atoms in the rod were allowed to relax with their positions determined by MD. The temperature was kept constant by velocity rescaling [6]. The forces used were those from the TTAM potential, BKS potential, or Born-Oppenheimer QM calculations at each time step. The stress was calculated by taking the sum of the forces parallel to the loading direction on the constrained atoms divided by the projected cross-sectional area of the nuclei that comprise the end caps. The cross-sectional area was updated after every step to account for any motion in x or y direction. The simulation was done at a temperature of 10K with a strain rate of 25 m/s and time steps of 2 fs. It is seen that for strains up to 4%, the Young's modulus (Y) given by the initial slope for the TTAM rod (Y = 1214) differs from that of the TH rod (Y=1026) by 18% while the difference between the BKS (Y=1516) and TH rod is even more than 50%. Clearly neither potential is a good candidate for an internally consistent multi-scale modeling in this system.

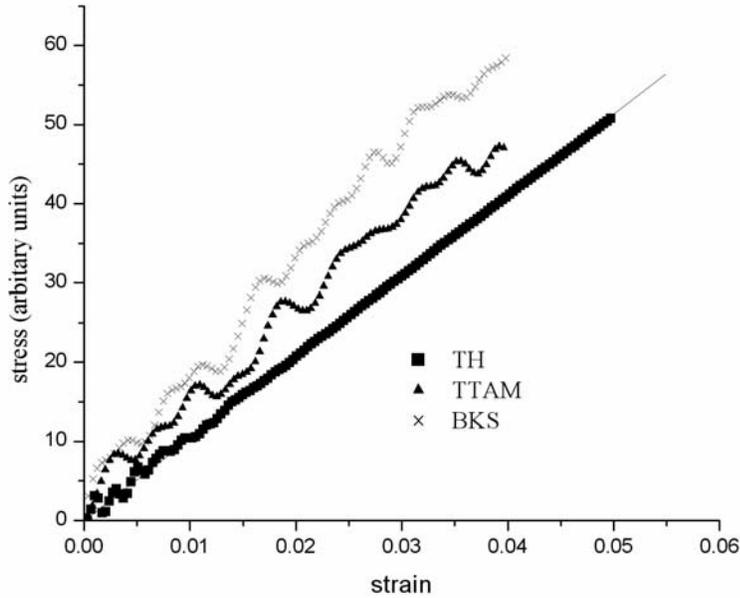

Figure 2. Comparison of stress-strain curves for standard pair potentials with that of TH

## II. Proposed Method and Objectives

The above results confirm that existing potentials may not be appropriate for multi-scale modeling of desired properties of specific systems of interest. Instead, it is proposed here that each development of a multi-scaling algorithm should include the construction of its own classical potential. That construction should be governed primarily by two components: the chosen quantum mechanical method used in the QM domain, and the particular physical properties requiring the QM treatment. In addition, it is required that there be a limiting domain (usually, near equilibrium) for which the CM and QM calculations of these properties are indistinguishable.

The proposed method is as follows. A QM method is chosen to represent the essential physical mechanisms not explicitly accounted for in a CM representation. Selected properties of the system are then calculated directly using the QM method. Here, these properties are the forces on nuclei incorporating non-local electronic configuration and exchange effects, both for equilibrium and weakly strained conditions. Next, a functional form for the potential energy is chosen for classical point particles replacing the atomic constituents. The parameters of the potential energy function are determined by some optimization method based on fitting the data calculated from the QM method.

There are several difficulties with the implementation of this general approach that must be addressed. Obtaining the *ab initio* QM data to constrain the potential can be problematic since the complexity of that calculation is indeed the motivation for multi-scale modeling. Consequently, the determination of the potential must be done on sufficiently small samples that limited benchmark QM calculations are feasible. The nanorod considered here satisfies this requirement for the quantum transfer Hamiltonian method, but is somewhat more difficult for the density functional method. Further comment on this is given below. A second issue is the choice of potential energy function. For practical MD simulation it is useful to restrict attention to two and three body potentials, but the functional form of these few body potentials is still optional. The functional forms that are frequently considered have their genesis in the theory of intermolecular interactions. However, their application to materials simulation is a rather speculative endeavor. In fact, one can argue that there is no fundamental origin for such potentials since the underlying electronic origin is both non local and many-body in nature.

Instead, the potentials are view as constructs of the multi-scale model "trained" to give selected material properties but with no unique status otherwise. Here, for simplicity, only pair potentials are considered and their functional forms are taken to be the same as the phenomenological forms currently in use for silica but with undetermined parameters.

Finally, the fitting of potential parameters to the QM data is neither straightforward nor unique since the parameter space is so large (here, 10 dimensional). Consequently, a variational method to minimize the difference between the CM and QM properties may encounter multiple local minima with different levels of stability for similar fitting of the data. Some compromise between a global search method and an effective local search is required in practice. This is accomplished here by the initial use of a genetic algorithm, followed by an over-all spatial scaling.

The objective here is to demonstrate the effectiveness of this approach to constructing a classical potential for the nanorod of Figure 1 as a test system. Specifically, it is required that the potential obtained should yield the correct equilibrium structure (bond lengths, bond angles) and small strain elastic properties, as given by the chosen quantum theory. The results are quite good and are shown to be robust with respect to the choice of quantum method used.

## III.   Quantum Methods and Potential Energy Form

The multi-scale problem of interest is to describe the nanorod strained uniaxially to failure. It is expected that bonds strained beyond a few percent can no longer be described accurately by a CM model. Under such conditions, the details of the shared electronic structure become important and the actual breaking of bonds corresponds to a complex charge rearrangement that can be accounted for only by a suitably accurate quantum theory. Two such quantum theories are considered here. First, and throughout the next two sections a transfer Hamiltonian (TH) method is used. This is a mean field, or Hartree-Fock like, description whose structure is simple and does not imply correlations. Consequently the solutions are Slater determinants with single electron wavefunctions determined quickly and analytically using a minimal basis set. The accuracy of this method is improved by parameterizing the Hamiltonian to fit certain predictions of the coupled cluster theory [7, 5] with single and double excitations (CCSD) where electron correlation is explicitly included. In this way, information about electron correlation is "transferred" to the mean field Hamiltonian. The primary advantage of this method is that practical calculations are possible on systems involving 100 – 200 nuclei.

The second choice for the QM method considered here, and discussed in Section VB, is a density functional theory (DFT) calculation. A Parallel multi-scale program package, based on Barnett's Born-Oppenheimer molecular dynamics BOMD [8] is used. A plane wave basis set and the Perdew-Burke-Ernzerhof (PBE) exchange-correlation functional [9] have been chosen for this calculation. This is a fully self-consistent calculation of the electronic structure including approximate, but explicit, electronic correlation. In this respect it might be considered superior to the TH method, but in fact there is no benchmark comparison of both with a more reliable coupled cluster calculation. However, it is important to emphasize at this point that the relative accuracy of the QM method is not at issue for the study here. The problem posed is: given a QM method, can a classical potential be constructed to reproduce the structure and elastic properties of that method? It is shown below that the answer to this question is affirmative, even though the TH and DFT nanorods have significant differences.

The classical potential energy function is assumed to be pairwise additive, with the pair potentials having the same functional form as TTAM [3] and BKS [4] potentials. Despite its known limitations, this form is chosen because of its simplicity in implementation and widespread use

$$U = \sum_{i<j} V_{ij}(r), \qquad V_{ij}(r) = \frac{q_i q_j}{r_{ij}} + a_{ij}\exp(-b_{ij}r_{ij}) - \frac{c_{ij}}{r_{ij}^6} \qquad (1)$$

Here *i,j* range over all Si and O ions of the rod. The first term of $V_{ij}$ is the Coulomb interaction for ions of charge $q_i$ and $q_j$. The remaining two terms are collectively called the "Buckingham" term. They model the short-range repulsive and dipole dispersion, or van der Waals, interactions, respectively. The Coulomb term is determined by a single parameter, either the charge on the silicon ion or on the oxygen ion, since there must be charge neutrality for the $SiO_2$ molecule (i.e., $2q_O = -q_{Si}$). There are altogether 10 parameters, consisting of the charge $q_{Si}$ and three pair parameters $a_{ij}, b_{ij}$ and $c_{ij}$ for each pair of interactions (O-O, Si-O and Si-Si).

## IV. Determination of Parameters

In the literature the parameters for the TTAM and BKS potentials have been chosen by fitting *ab initio* Hartree-Fock potential energy surfaces of $SiO_4(4-)+4e^+$ and $SiO_4H_4$ respectively. Furthermore, the Si-Si interactions in BKS potential are arbitrarily set equal to zero. The resulting values are given in Table 2 below. It is noted that this fitting was done with a low level Hartree-Fock approximation and the surface was explored at or near equilibrium regions only. In contrast, the parameters are determined here by the method described above. Since MD requires only the forces between ions, these forces determined directly from the quantum mechanics are used as primary reference data. Ideally, these forces should be zero for the correct equilibrium structure and its stability. Furthermore, to assure good elastic properties information about these forces for small strained conditions also is used in the search for appropriate parameters.

The standard approach for parameterization consists of following steps: 1) choose properties of interest (total forces on atoms), 2) choose reference data (values of forces for equilibrium and strained configurations), 3) perform some estimation of the parameters, 4) test against reference data with some error function, 5) choose a variational method for the next estimate and convergence of the procedure.

In our case the reference quantities are the QM forces on atoms of the rod in equilibrium, and when strained from its equilibrium structure. The error function *L*, for testing the quality of any chosen set of parameters is the weighted sum of absolute difference between the calculated forces from the trial potential and the reference forces, denoted by *g(i)* and *f(i)* respectively for the forces on the $i^{th}$ ion of the rod

$$L = \sum_k a_k \sum_i |g_k(i) - f_k(i)| \qquad (2)$$

The index *i* runs over all ions of the rod, while *k* denotes the sum over data sets for different structural configurations of the rod (different amounts of strain). The positive numbers $a_k$ are chosen weights assigned to the different strain sets with $\sum_k a_k = 1$. In general the greatest weight was given to the equilibrium structure, with the weight decreasing for data with increasing strain. A typical distribution of weights was $a_k = (0.5, 0.2, 0.1, 0.1, 0.05, 0.05)$ for percent strains (0, 1, 2, 3, 4, 5), respectively. Other forms for *L* (e.g., root mean square of differences or normalizing the difference) gave similar results.

This error function is minimized in a space whose dimension is the number of adjustable parameters. The problem of such a high dimensional domain is the existence of several local minima and maxima. Gradient optimizing routines [10] seek extrema in the neighborhood of initial chosen point without sampling the whole space and hence are only local in scope. Further improvement can be sought only through some sort of random restart. One can repeatedly restart the search algorithm at other randomly chosen points but as the dimension of the space increases such a process becomes highly inefficient. Therefore a genetic algorithm [11] has been used which doesn't suffer from such disadvantages. Furthermore, the genetic algorithm (GA) doesn't require calculation of derivatives and hence is more efficient. The standard PIKAIA code [12] for the GA has been used. However once a neighborhood of an extrema is reached the GA converges slowly near each extremum since they rely primarily on the mutation rate to generate small incremental changes in the population. If the global minimum has not been located the search will begin to stagnate. This is due to the successive removal of comparatively less fit individuals from the population, which, after many generations, results in a population of parameter sets which are all more or less equally fit and the evolution of the parameters stalls. This problem is analogous to that of the gradient methods getting trapped in local minima.

For a given QM data set the parameters obtained by the GA assures forces accurate to the chosen tolerance. However, this does not assure that the equilibrium structure obtained using these parameters is stable. Stability of the rod was tested checked using a Broyden-Fletcher-Goldfarb-Shanno [13] energy minimization simulation at 0 K temperature. In this method [13], the function to be minimized (energy in our case) is expanded to second order in terms of parameter set p (different geometries of the rod in our case) as

$$f(p) \approx f(p_o) + g^T \times (p - p_o) + \frac{1}{2}(p - p_o)^T \times H \times (p - p_o) \qquad (3)$$

where $g$ is the gradient of the function and $H$ is the Hessian. The derivative of the function is calculated numerically. In order to have the geometry with minimum energy requires the determinant of Hessian to be positive. The optimization proceeds with the generation of a new geometry $p$ given an old geometry $p_o$ using $(p - p_o) = H^{-1}g$. This process continues until $f$ goes below some acceptable tolerance. As anticipated, it was found that there are numerous local minima in the parameter space, and several possible sets of potential parameters were found from the GA that yielded a stable rod.

Rather than require convergence of the GA to a higher level of accuracy for the forces, a lower tolerance was used to get a first estimate of the parameters. From the resulting set yielding a stable rod, the one with least structural error in bond lengths and bond angles was chosen for the second stage of analysis. At that point one might think of switching to Newton methods, i.e. to construct hybrid schemes of GA and gradient based methods using GA results as a starting guess. Instead it was noticed that the structure resulting from parameters of the first estimate differed from the QM rod mainly by an overall spatial scaling. Consequently, the final parameter set was obtained by a simple rescaling of those from the first estimate (see Appendix), under the constraints of the given QM structure and linear stability.

## V. Results

### A. Transfer Hamiltonian Quantum Mechanics

Consider now the case for which the TH QM method is used to generate forces for the entire rod, under conditions of equilibrium and strains up to 10%. Initially, only equilibrium and small strain data (up to 5%) were used in the GA error function L. Together with subsequent scaling a set of potential parameters yielding good structure and elastic properties up to 5% was obtained, as desired. As expected, deviations in the elastic properties for the TH and the new potential were observed at still higher strains. Unfortunately, the classical rod based on the new potential became more ductile than the quantum rod. Although this is the domain for which the classical potential no longer represents the quantum rod, it is desirable for multi-scale modeling that CM domain should grow *less* ductile than the QM domain so that the large strain behavior is localized in the QM domain where it is properly treated. This qualitative property was attained by repeating the parameterization using QM force data up to 10% expansion in the GA. The data for forces at the end caps were excluded in this case as the bonding at large strain becomes significantly different from that in the bulk and is misleading the GA. (It is observed that the information about the bulk force data is sufficient to generate the end cap structure.)

Table 1 shows the final parameters for the CM potential constructed in this way, along with the standard TTAM and BKS parameters.

| Parameters | TTAM | BKS | New - Classical - TH potential |
|---|---|---|---|
| $q_{Si}$ | 2.4 | 2.4 | **2.05** |
| $q_O$ | -1.2 | -1.2 | **-1.025** |
| $a_{oo}$ | 1756.98 | 1388.773 | **39439.87** |
| $b_{oo}$ | 2.846 | 2.760 | **4.66** |
| $c_{oo}$ | 214.75 | 175.00 | **5.85** |
| $a_{osi}$ | 10722.23 | 18003.757 | **240101.9** |
| $b_{osi}$ | 4.796 | 4.873 | **7.88** |
| $c_{osi}$ | 70.739 | 133.538 | **2.06** |
| $a_{sisi}$ | 8.73E+08 | 0 | **27530.45** |
| $b_{sisi}$ | 15.22 | 0 | **21.42** |
| $c_{sisi}$ | 23.265 | 0 | **1.55** |

* The units of $a_{ij}$, $b_{ij}$ and $c_{ij}$ are eV, $(A^o)^{-1}$ and eV-$(A^o)^6$ respectively

Table 1. Parameters for New Potential

It is noted that the values of the charges and $b_{ij}$ for the new potential are similar to those of TTAM or BKS, but that the $c_{ij}$ are always much smaller for the new potential. This suggests a very much more repulsive character of this potential at short-range, as illustrated in Figure 3 for the Si-O interactions.

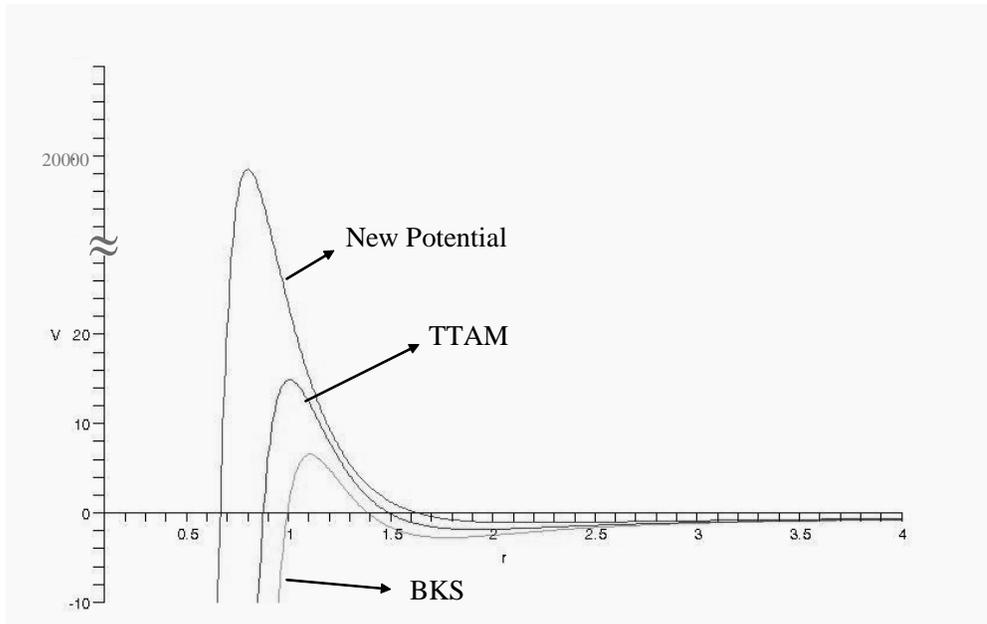

Figure 3. Si-O Interactions for the three potentials

Table 2 shows the comparison of a several bond-lengths at equilibrium for the three potentials and those for the reference TH rod. The percent errors for the new potential are seen to be considerably smaller than those of the TTAM or the BKS potentials, indicating a good agreement between the CM and TH rods at equilibrium.

| Bond lengths and Angles | TH | New Cl-TH potential | % error | TTAM | % error | BKS | % error |
|---|---|---|---|---|---|---|---|
| *In Silica Planes* | | | | | | | |
| Si-O | 1.641 | 1.642 | **0.04** | 1.65 | 0.76 | 1.611 | 1..5 |
| <Si-O-Si | 170.06 | 173.7 | **2.14** | 162.9 | 4.2 | 161.1 | 5.2 |
| *Between Planes* | | | | | | | |
| <O-Si-O | 103.8 | 104.3 | **0.4** | 104.6 | 0.7 | 104.9 | 1.04 |
| *End Caps* | | | | | | | |
| Si-O | 1.71 | 1.67 | **2.5** | 1.67 | 2.5 | 1.62 | 5.1 |
| <Si-O-Si | 102.03 | 103.2 | **1.14** | 100.6 | 1.45 | 100.8 | 1.23 |
| Length | 16.49 | 16.34 | **0.9** | 16.31 | 1.1 | 15.86 | 3.8 |
| Diameter | 6.55 | 6.54 | **0.15** | 6.57 | 0.35 | 6.41 | 2.0 |

Table 2. Comparison of structure of the rod with the different potentials

This new potential next was used in MD simulations to obtain the stress-strain curves for a classical model of the nanorod. The simulation was carried out under the same conditions as for the TH (temperature of 10 K and pulling the end-caps at 25 m/s). Figure 4 shows the results for this new potential, together with those for the TTAM and BKS potentials, in comparison with the TH results. The stress-strain curve obtained from the new potential agrees well with the TH curve up to about 5 % strain. Accordingly, the Young's modulus obtained from the new potential agrees very well with that from the reference TH. At larger strains the CM results deviate from the TH towards larger stress, as desired. The oscillations in the curves for the classical potentials were diminished considerably by performing MD simulations at lower temperatures (down to 0.2 K) and by adiabatic expansions (0 K). The Young's modulus does not change significantly in the range 0 K < T < 50 K.

For completeness, a comparison of the CM and TH stress-strain curves up to strains beyond failure are shown in Figure 5. Of course, the CM potentials no longer resemble the QM material in the large strain domain. However, it illustrates clearly that the new potential satisfies all the requirements for a multi-scale modeling of the QM rod: accurate near equilibrium properties and less ductile elastic properties far from equilibrium.

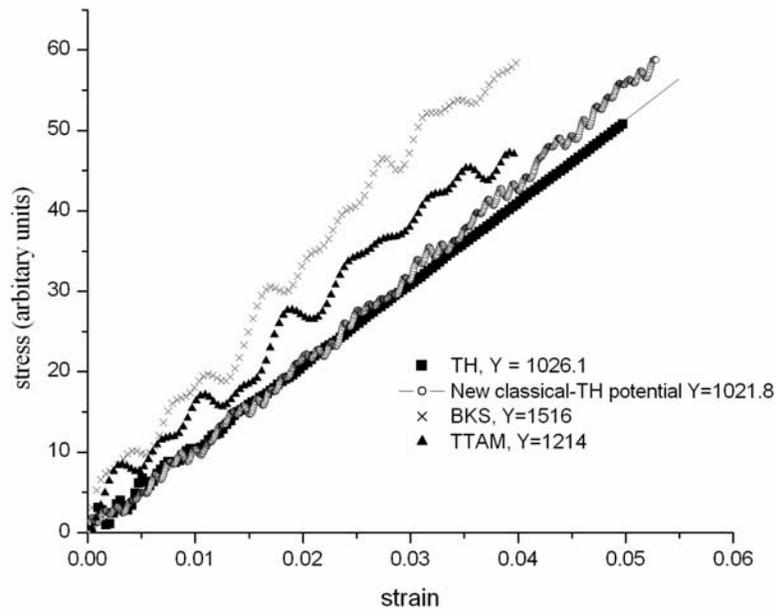

Figure 4. Stress-strain curves for different potentials

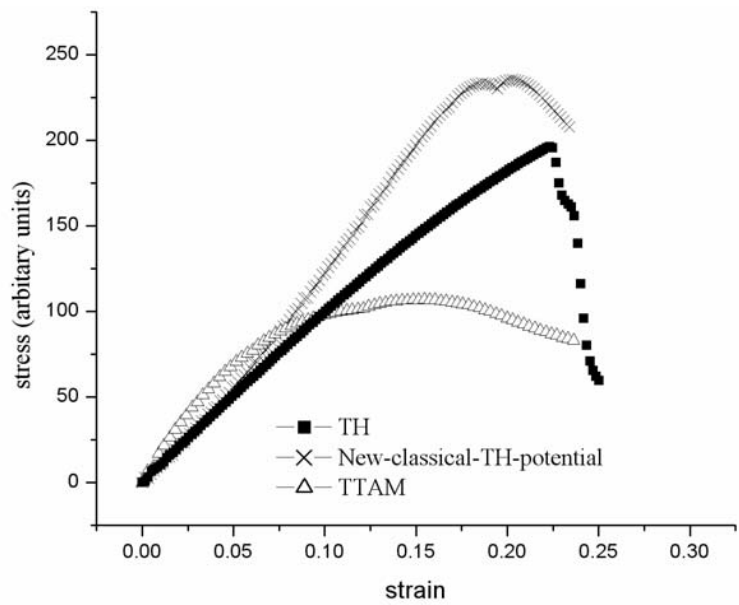

Figure 5. Behavior of the stress-strain curves near fracture

## B. Density Functional Theory Quantum Mechanics

The same method for determining a CM potential was tested for the case of an underlying quantum mechanics determined from the density functional theory described above. In this case the QM calculations are considerably more time intensive so that only equilibrium and selected adiabatic strain configurations were calculated for the QM. The equilibrium structure was determined by sequential DFT calculations and nuclear relaxation to find the minimum energy configuration. The strained configurations were obtained from an affine transformation of the minimum energy configuration by 1, 2, 3, and 4 %, with a single DFT calculation of forces at each of the expanded configurations. The potential parameters were determined using this data in the GA, followed by scaling. Table 3 shows the resulting parameters in comparison to the TTAM and BKS parameters. Again, we see that the charge on the ions is lower, in this case almost halved, from that found in the TTAM and BKS potentials. Also, the van der Waals interaction, the $c_{ij}$'s, are much less than in the published potentials.

| Parameters | New - Classical –DFT potential | BKS | TTAM |
|---|---|---|---|
| $q_{Si}$ | 1.4 | 2.4 | 2.4 |
| $q_O$ | -0.7 | -1.2 | -1.2 |
| $a_{oo}$ | 1281.153 | 1388.773 | 1756.98 |
| $b_{oo}$ | 3.654891 | 2.760 | 2.846 |
| $c_{oo}$ | 3.899369 | 175.00 | 214.75 |
| $a_{osi}$ | 7846.292 | 18003.757 | 10722.23 |
| $b_{osi}$ | 6.11894 | 4.873 | 4.796 |
| $c_{osi}$ | 1.294512 | 133.538 | 70.739 |
| $a_{sisi}$ | 6.28E+08 | 0 | 8.73E+08 |
| $b_{sisi}$ | 25.41433 | 0 | 15.22 |
| $c_{sisi}$ | 0.595124 | 0 | 23.265 |

Table 3. Comparison of DFT Potential Parameters

Table 4 shows a comparison of the equilibrium structure. The agreement between the CM and DFT rods is quite good. Similarly, the elastic properties also are in quite good agreement as shown in Figure 6. Also shown is the result for the TTAM potential. All curves in Figure 6 were calculated for the same adiabatic expanded configurations.

| Bond lengths and Angles | DFT | New DFT Classical | % error |
|---|---|---|---|
| *In Silica Planes* | | | |
| Si-O | 1.62 | 1.62 | |
| <Si-O-Si | 158.02 | 157.96 | 0.04 |
| *Between Planes* | | | |
| <O-Si-O | 107.15 | 103.74 | 3.0 |
| *End Caps* | | | |
| Si-O | 1.64 | 1.63 | 0.6 |
| <Si-O-Si | 92.57 | 100.5 | 8.0 |
| Length | 15.79 | 16.1 | 1.96 |
| Diameter | 6.44 | 6.44 | |

Table 4. Comparison of Structure obtained from the New DFT potential

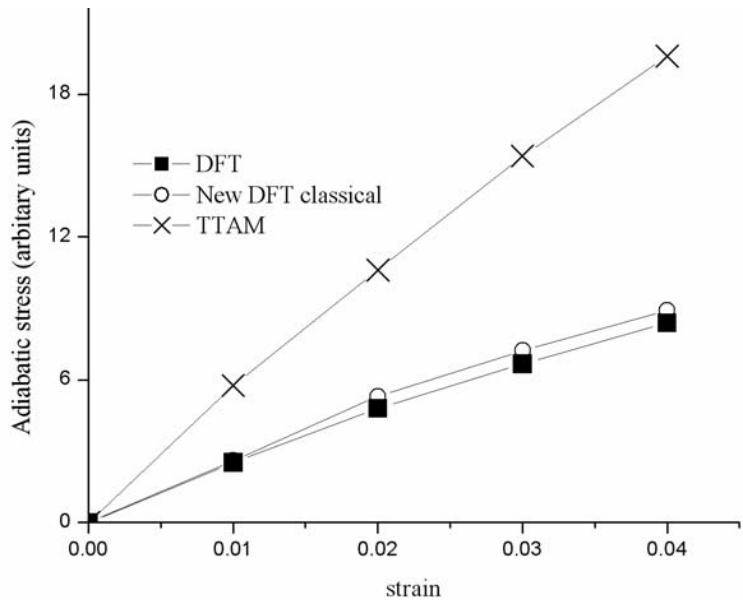

Figure 6. Comparison of adiabatic stress curves for DFT, new-DFT-classical and TTAM potential

## III. Conclusions

The use of *ab initio* QM data to determine a corresponding CM potential for multi-scale modeling has been illustrated in detail here for a non-trivial mesoscopic system, the $SiO_2$ nanorod. The objective of reproducing equilibrium structure and linear elastic properties has been attained to quite good accuracy. Furthermore, it has been demonstrated that the method is not sensitive to the choice for the fundamental QM method used. The quality of the CM potential obtained for both the TH method and the DFT method is comparable in each case.

Although the approach proposed here is to tailor the potential to each multi-scale modeling problem considered, it is tempting to ask to what extent a potential constructed for the nanorod is representative of other silica based systems. This kind of extrapolation must be used with some care, and is not in the constructive spirit of the presentation here. Nevertheless, for practical purposes it appears that the nanorod potentials do have a wider applicability. Two examples have been explored with positive preliminary results. The TH CM potential has been used for MD simulation of bulk silica glass properties, such as thermodynamics (density) and structure (radial distribution functions) with quite satisfactory results in comparison with experimental data [14]. The second example is a ten membered $SiO_2$ nanoring [15], where equilibrium structure (bond angles) obtained using the CM potential in MD simulation gives good agreement with QM results based on the TH method. These results are somewhat surprising since the local structure of the glass and nanoring are quite different from that of the rod. This suggests a new phenomenology whereby *ab initio* data for a specific system and specific properties is used to construct a classical potential, followed by an empirical exploration of its range of validity elsewhere.

The study here posed the problem of constructing a CM potential to represent the properties of a given approximate QM description of a system. This is separate from the question of how accurate that approximate QM description may be. Consequently, in the discussions above we have compared the CM potential obtained from TH data with the results of the TH QM structure and elasticity. Similarly the CM potential obtained from DFT data has been compared to results of DFT QM. In closing, however, it is appropriate to note that aside from any issues of constructing classical potentials the calculations performed here at the QM level lead to noticeably differences between the TH and DFT quantum rods. For example the TH rod is about 4% longer and 2% wider than the DFT rod; the Si-O-Si bond angle in plane is about 7% larger for the TH rod. While these are not very large differences they lead to quite different elastic properties as shown in Figure 7. This puts in context the task of constructing CM potentials representing a given QM method. Generally, multi-scale modeling requires an efficient *ab initio* method to describe a mechanism (e.g. charge transfer) but whose absolute accuracy is not known. Consequently, the accuracy of the CM representation is required only within this same tolerance.

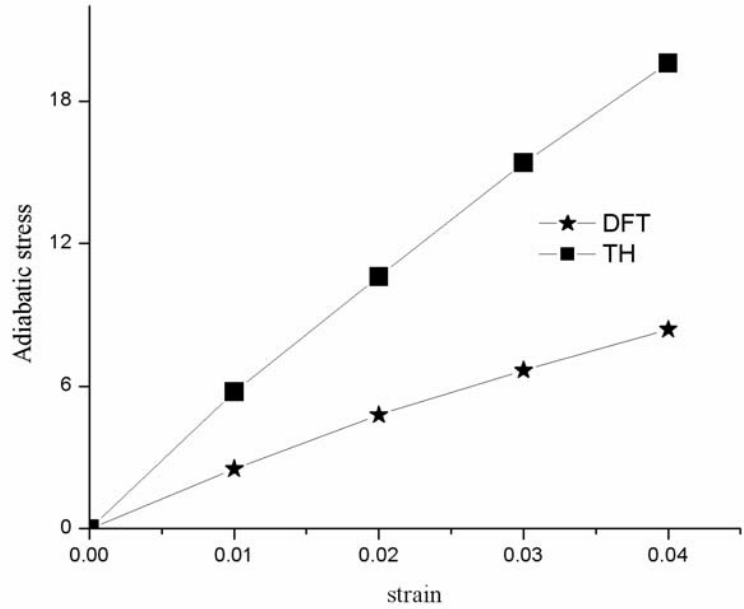

Figure 7. Figure showing the difference of elastic properties between the TH and DFT rods.

## V. Acknowledgements

This research was supported by NSF – ITR under Grant No. 0325553. The authors are indebted to Drs. D. E. Taylor, S. B. Trickey, P. Deymier, K. Muralidharan and L.Wang for helpful discussions and assistance.

APPENDIX – RESCALING METHOD

The parameters of Table 1 obtained from the GA represent stable equilibrium structures, but these structures still differ somewhat from those of the underlying quantum TH calculation. To improve the potential further new potential parameters are determined from these by "mapping" the GA structures onto the correct TH structure through an affine scaling of the coordinates. Let $p_{ij} = (q_i, q_j, a_{ij}, b_{ij}, c_{ij})$ denote one set of parameters from Table 2 and let $r_{ij}$ be the separation for the pair i,j at equilibrium. Also, let $r_{ij}'$ be the corresponding equilibrium separation from the TH. The objective is to find a new set of parameters $p'_{ij}$ for the pair potential such that the new equilibrium configuration is at $r_{ij}'$.

As a first step require that the pair forces under change of parameters and rescaling of the separation are proportional to the original forces

$$\mathbf{F}_{ij}(r_{ij}', p_{ij}') \propto \mathbf{F}_{ij}(r_{ij}, p_{ij})$$

The sum of the forces $\mathbf{F}_{ij}(r_{ij}, p_{ij})$ over i for each j is (nearly) zero as a result of the GA search. This condition assures that the new parameters will yield a new equilibrium at $r_{ij}'$ in agreement with the TH calculation. This is strictly true only if the $r_{ij}$ and $r_{ij}'$ are related by the same scale transformation for every pair, which is observed to be approximately the case. Next, the stability of this new equilibrium condition is assured by the second constraint

$$\frac{d}{dr_{ij}'}\mathbf{F}_{ij}(r_{ij}', p_{ij}') = \frac{d}{dr_{ij}}\mathbf{F}_{ij}(r_{ij}, p_{ij})$$

For the chosen pair potential

$$V_{ij}(r) = \frac{q_i q_j}{r_{ij}} + a_{ij}\exp(-b_{ij} r_{ij}) - \frac{c_{ij}}{r_{ij}^6}$$

the magnitude of the force and its derivative are

$$F_{ij}(r) = -\frac{\partial V_{ij}}{\partial r_{ij}} = \frac{q_i q_j}{r_{ij}^2} + a_{ij} b_{ij} \exp(-b_{ij} r_{ij}) + \frac{6c_{ij}}{r_{ij}^7}$$

$$\frac{\partial F_{ij}}{\partial r_{ij}} = -2\frac{q_i q_j}{r_{ij}^3} - a_{ij} b_{ij}^2 \exp(-b_{ij} r_{ij}) - \frac{42c_{ij}}{r_{ij}^8}$$

Equating the force derivatives for the original and rescaled potentials gives

$$q_i'q_j' = \left(\frac{r_{ij}'}{r_{ij}}\right)^3 q_i q_j, \quad a_{ij}'b_{ij}'^2 \exp(-b_{ij}'r_{ij}') = a_{ij}b_{ij}^2 \exp(-b_{ij}r_{ij}), \quad c_{ij}' = \left(\frac{r_{ij}'}{r_{ij}}\right)^8 c_{ij},$$

while the condition for the forces to be proportional gives in addition $b_{ij}' = b_{ij}(r_{ij}/r_{ij}')$.